\documentclass[sigconf, nonacm]{acmart}

\AtBeginDocument{%
  }

\begin{document}

\title{High Precision Audience Expansion via Extreme Classification in a Two-Sided Marketplace}

\author{Dillon Davis}
\email{dillon.davis@airbnb.com}
\affiliation{%
  \institution{Relevance}
  \institution{Airbnb, Inc.}
  \city{San Francisco, CA}
  \country{USA}
}
\author{Huiji Gao}
\email{huiji.gao@airbnb.com}
\affiliation{%
  \institution{Relevance}
  \institution{Airbnb, Inc.}
  \city{San Francisco, CA}
  \country{USA}
}
\author{Thomas Legrand}
\email{thomas.legrand@airbnb.com}
\affiliation{%
  \institution{Relevance}
  \institution{Airbnb, Inc.}
  \city{San Francisco, CA}
  \country{USA}
}

\author{Juan Manuel Caicedo Carvajal}
\email{juan.caicedocarvajal@airbnb.com}
\affiliation{%
  \institution{Relevance}
  \institution{Airbnb, Inc.}
  \city{San Francisco, CA}
  \country{USA}
}

\author{Malay Haldar}
\email{malay.haldar@airbnb.com}
\affiliation{%
  \institution{Relevance}
  \institution{Airbnb, Inc.}
  \city{San Francisco, CA}
  \country{USA}
}

\author{Kedar Bellare}
\email{kedar.bellare@airbnb.com}
\affiliation{%
  \institution{Relevance}
  \institution{Airbnb, Inc.}
  \city{San Francisco, CA}
  \country{USA}
}

\author{Moutupsi Paul}
\email{moutupsi.paul@airbnb.com}
\affiliation{%
  \institution{Search}
  \institution{Airbnb, Inc.}
  \city{San Francisco, CA}
  \country{USA}
}

\author{Soumyadip Banerjee}
\email{soumyadip.banerjee@airbnb.com}
\affiliation{%
  \institution{Search}
  \institution{Airbnb, Inc.}
  \city{San Francisco, CA}
  \country{USA}
}

\author{Liwei He}
\email{liwei.he@airbnb.com}
\affiliation{%
  \institution{Relevance}
  \institution{Airbnb, Inc.}
  \city{San Francisco, CA}
  \country{USA}
}

\author{Stephanie Moyerman}
\email{stephanie.moyerman@airbnb.com}
\affiliation{%
  \institution{Relevance}
  \institution{Airbnb, Inc.}
  \city{San Francisco, CA}
  \country{USA}
}

\author{Sanjeev Katariya}
\email{sanjeev.katariya@airbnb.com}
\affiliation{%
  \institution{Relevance}
  \institution{Airbnb, Inc.}
  \city{San Francisco, CA}
  \country{USA}
}

\renewcommand{\shortauthors}{Davis et al.}

\begin{abstract}
Airbnb search must balance a worldwide, highly varied supply of homes with guests whose location, amenity, style, and price expectations differ widely. Meeting those expectations hinges on an efficient retrieval stage that surfaces only the listings a guest might realistically book, before resource intensive ranking models are applied to determine the best results. Unlike many recommendation engines, our system faces a distinctive challenge — location retrieval — that sits upstream of ranking and determines which geographic areas are queried in order to filter inventory to a candidate set. The preexisting approach employs a deep bayesian bandit based system to predict a rectangular retrieval bounds area that can be used for filtering. The purpose of this paper is to demonstrate the methodology, challenges, and impact of rearchitecting search to retrieve from the subset of most bookable high precision rectangular map cells defined by dividing the world into 25M uniform cells.
\end{abstract}


\maketitle

\section{Introduction}

\null \quad Airbnb has grown into a truly global marketplace: millions of unique homes span bustling city centres, quiet suburbs, rural countrysides, and every kind of coastline in between.  Guests arrive with highly-specific wishes – particular neighbourhood vibes, family-friendly amenities, distinctive architectural styles, strict budgets – and they expect those preferences to be respected no matter where they search.

\null \quad Traditional lodging is usually clustered in predictable tourist hubs, but Airbnb's supply is often dispersed well outside those cores.  Consider a scenario where a family is planning a visit to a city that restricts short term-rentals and reached hotel capacity - there may be spacious homes just across the municipal border that could be a better fit for this family. Our challenge is to make such hidden-gem inventory discoverable \emph{immediately}, without forcing guests to pan and zoom for minutes on the map.

\null \quad A search session almost always starts with a typed destination in the search bar.  The system must produce a ranked list of "most likely to be booked" listings within a few hundred milliseconds.  Roughly half of guests later move the map, but the other half never look beyond that first viewport.  Showing the right homes in the very first response is therefore critical for both guest satisfaction and host success.

\null \quad Scale compounds the problem: Airbnb hosts more than 8 million active listings and processes over 400 million nights and experiences per year.  Ranking every listing for every query is impossible, and filtering strictly to the named destination discards valuable choices in adjacent areas. The search pipeline therefore begins with \textbf{location retrieval} – a step that selects a geographically constrained subset of listings that a guest \emph{might plausibly book}.

\begin{figure}[h]
  \centering
  \includegraphics[width=1\linewidth]{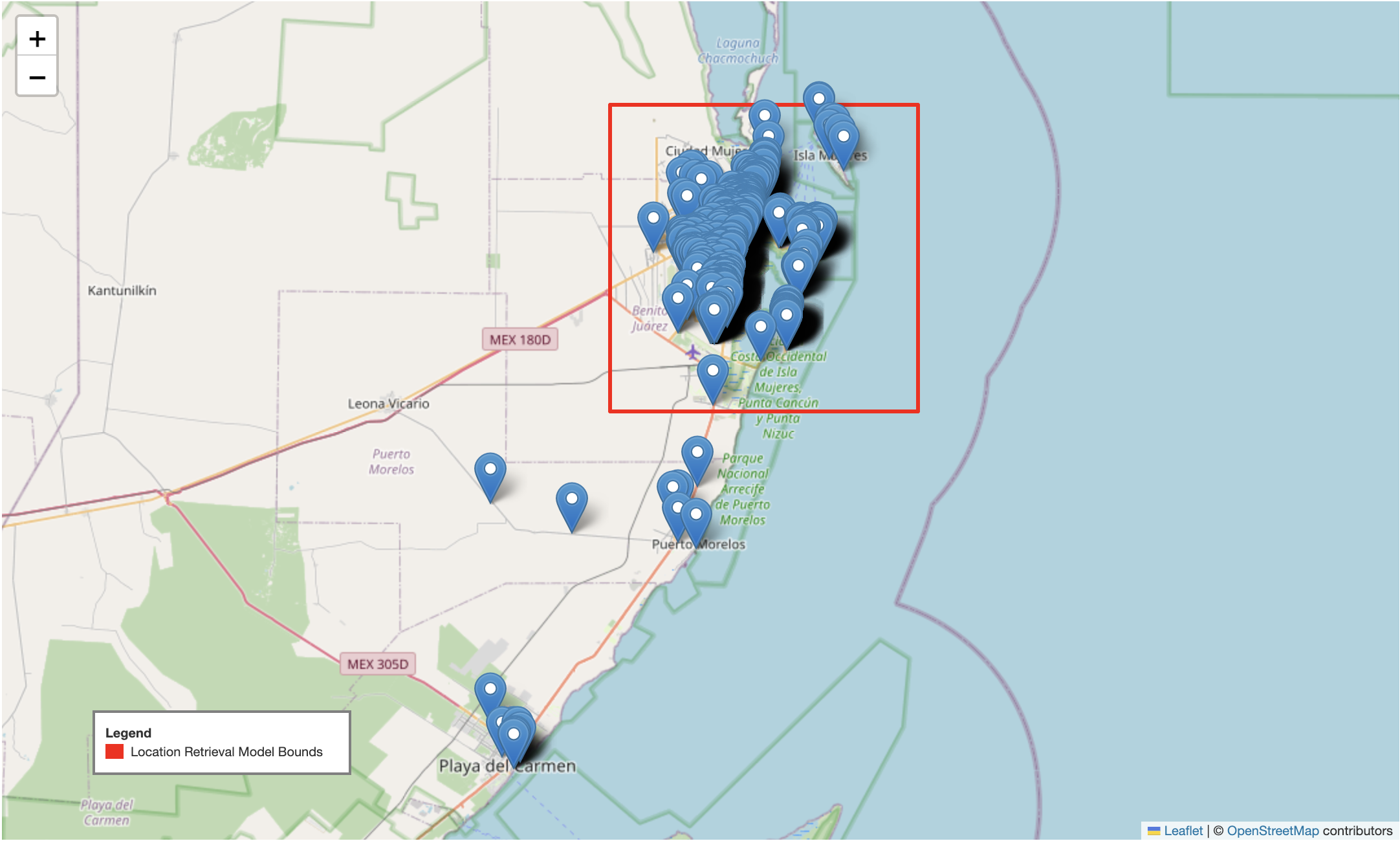}
  \caption{Locations of Bookings from Cancun Searchers and Existing Retrieval Bounds}
  \label{cancun_bookings}
\end{figure}

\null    \quad The existing deep bayesian bandit based system \cite{dillontransforming2024} infers relevant retrieval bounds using a model trained on search queries with booked listing location labels. It optimizes a weighted piecewise loss with multiple objectives such as including locations of relevant booked listings and minimizing the size of the retrieval bounds. While performant, this rectangular retrieval bound approach is inherently unable to precisely improve recall of bookable locations without including many other unbookable areas. For example, Figure \ref{cancun_bookings} demonstrates the existing model based retrieval bounds and booked listing locations shown as pins for Cancun searches . There are intermittent beach destinations along the shoreline that are popular booking destinations for Cancun searchers like Playa del Carmen. However, the retrieval bounds model would have to include many unbooked destinations in between Cancun and Playa del Carmen in the retrieval set in order to expand to include Playa Del Carmen.

    \quad Our existing location retrieval bounds solution often surfaces unbooked locations as well due the same limitations mentioned above. For example, ~39.8\% of results in candidate retrieval sets for "San Francisco" searches are in locations that were not previously booked for "San Francisco" searches. Furthermore, ~14.7\% of "San Francisco" search results shown to users are in locations that were not previously booked for "San Francisco" searches.

\begin{figure}[h]
  \centering
  \includegraphics[width=1\linewidth]{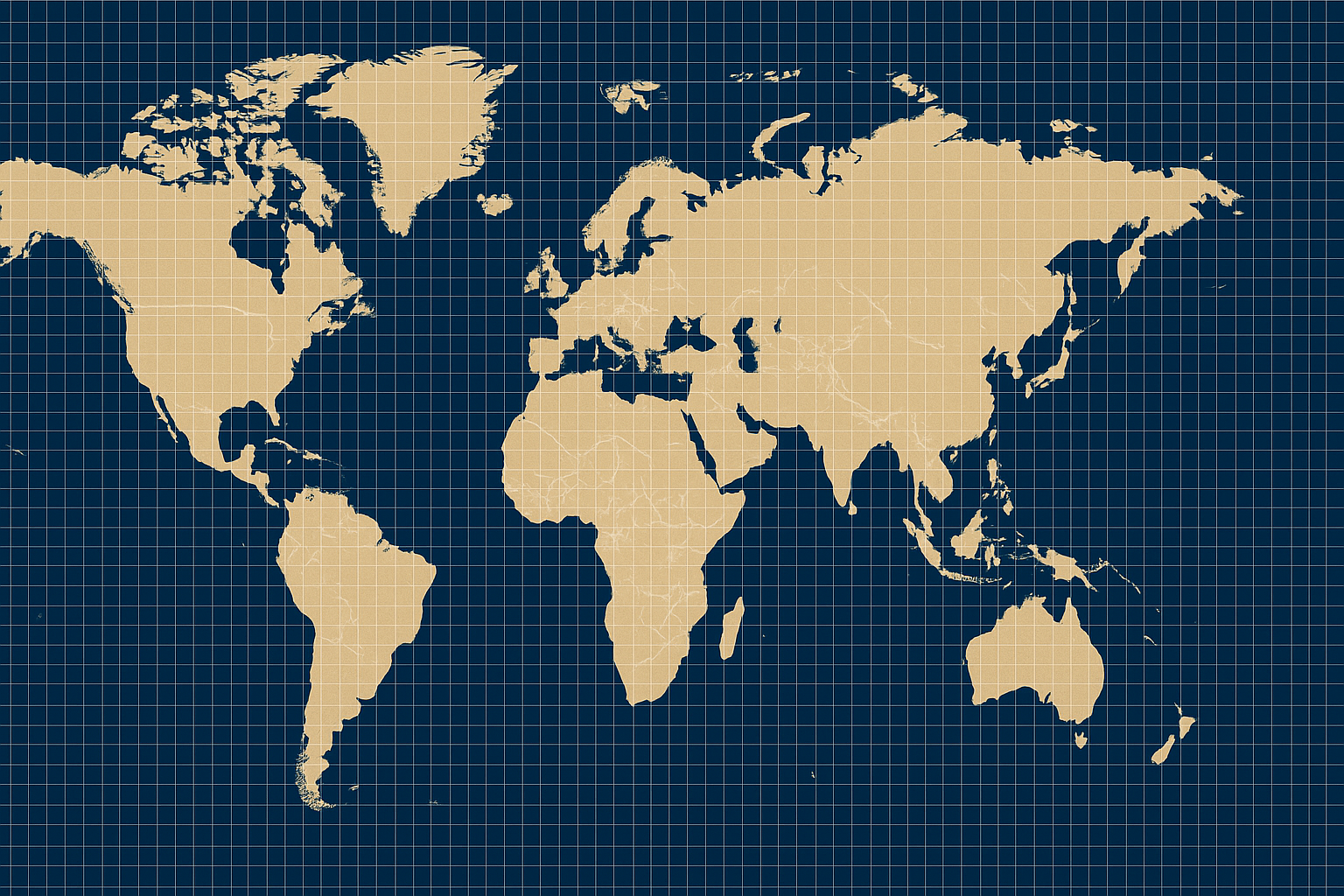}
  \caption{World Map split into S2 Cells}
  \label{s2cellworld}
\end{figure}

\null \quad Considering these limitations and evidence, we rearchitect location retrieval to retrieve listings from a set of relevant, high precision categorical location cells shown for the world in Figure \ref{s2cellworld} instead of one large contiguous rectangular retrieval bounds area. This new approach will allow us to simultaneously remove inventory in unbookable locations from our existing search results and surface new inventory from previously unsurfaced bookable locations.

\section{Related Work}
\label{sec:related}

\paragraph{Location retrieval in two-sided marketplaces.}
Commercial marketplaces typically narrow the candidate pool with simple geometric rules.  Foursquare's local-search stack, for example, expands a rectangular bounding box until a sufficient number of venues are found \cite{foursquare2010}.  Uber's surge-pricing and supply forecasts rely on the hexagonal \textsc{H3} index, but still aggregate demand at coarse grid levels \cite{uberh32018}. Airbnb's production system inferred rectangular retrieval bounds via heuristics and later a contextual bandit \cite{dillontransforming2024}. We advance this line by predicting the exact set of fine-grained S2 cells most likely to yield bookings, removing the shape assumptions of rectangles.

\paragraph{Spatial indexing and S2 geometry.}
The S2 library partitions the sphere into a quad-tree of equal-area cells and has been adopted for geospatial storage and querying~\cite{s22017}.  Prior work exploits S2 for \emph{indexing}; we are, to our knowledge, the first to treat individual level-11 cells as labels in a large-scale machine-learning model served online under sub-second latency.

\paragraph{Extreme multi-class classification.}
Predicting millions of labels has been explored in text domains~\cite{parabel2018,attentionxml2019,mip2016}. Tree-based approaches such as \textsc{Parabel}~\cite{parabel2018} partition the label space to avoid computing a full softmax, while embedding methods rely on nearest-neighbor search over learned prototypes~\cite{mip2016}. Our work adapts \emph{sampled softmax}, showing that a dense neural network can scale to 300K\,+ geospatial labels learned from sparse data without significant latency overhead.

\paragraph{Joint retrieval and ranking.}
Two-stage architectures that first retrieve a candidate set and then apply a heavier ranker are ubiquitous in web search and recommender systems~\cite{dillontransforming2024,haldar2023learning}. Recent studies show advanced techniques like deep two-tower models or reinforcement learning that optimize end-to-end metrics~\cite{haldar2020improving,tan2023optimizing} are impactful. Our solution plugs into the same two-stage pipeline but focuses on the
\emph{geospatial} dimension of retrieval, preserving compatibility with downstream rankers while increasing precision and supply coverage.

\vspace{0.5em}
\noindent
By bridging fine-grained spatial indexing with extreme classification techniques, this work contributes a new retrieval layer that demonstrably lifts bookings in a global two-sided marketplace.

\section{Approach}
\null \quad We start by formulating a multiclass classification problem. We construct a training dataset using searches and their attributed booked listings' locations and train a model to predict a discretized form of booked listing locations given the search context. At search time, we score the model using the search's associated features to construct a probability distribution over the predefined categorical space that discretizes continuous listing location coordinates. We then only retrieve listings from classes whose probability meets a certain threshold.

\subsection{Multiclass Classification}
\[
P\!\bigl(C = c_i \mid B = 1,\; Q = q\bigr), \qquad c_i \in \mathcal{S}^{(11)} .
\]

\begin{align*}
C                &\,=\, \text{random variable for the booked S2 cell} \\[2pt]
c_i              &\,=\, \text{a specific level-11 S2 cell label}                          \\[2pt]
B                &\,=\, \text{booking indicator}        \\[2pt]
Q                &\,=\, \text{random variable representing the search}      \\[2pt]
d                &\,=\, \text{concrete value of the search}                    \\[2pt]
\quad \mathcal{S}^{(11)} &\,=\, \text{set of all level-11 S2 cells}                             \\[2pt]
                 &      \text{(\(\lvert\mathcal{S}^{(11)}\rvert = 25{,}165{,}824\))}      
\end{align*}
Our training examples are constructed using Airbnb searches whose search results contain a listing that the searcher later booked. The labels are the locations of the search results that were later booked by the searcher. We map the latitude and longitude location coordinates of the booked search result to a predefined categorical space to construct a traditional multiclass classification problem.

One of the key considerations when constructing training data for location retrieval is the attribution of booked listings to searches and destinations. The Airbnb search to booking journey is very complex due to the high-value nature of transactions compared to traditional two-sided marketplace systems. Bookers typically issue $\sim$49 searches and view $\sim$9 listings before booking their desired listing. They typically begin their journey by searching for a destination with the search bar and then apply filters, pan or zoom the map, or search new destinations. Instead of the original search through the search bar, many guests find a suitable listing by panning and zooming the map to explore more listings and areas. Therefore, it's crucial to attribute booked listings found through subsequent map searches to the original searched destination in order to fully capture guests' location preferences for a given travel destination. For example, this is how we learn that Cancun searchers are willing to stay in Playa Del Carmen even though Playa Del Carmen listings do not appear in the initial search results for Cancun searches. However, this also introduces outliers into the dataset such as when someone searches for San Francisco, moves the map across the world, and books a listing for a completely different trip.

\subsection{Features}
We consider the following features to represent the user's search in order to predict a probability distribution over the categorical location space.
    \begin{enumerate}
        \item Unique id of the specific cell on the earth's surface containing the searched location's center of varying cell sizes
        \item Whether the searched location is an address, POI, street, neighborhood, or city
        \item Country of the searched location
        \item Request origin country
        \item Number of guests
        \item Whether the search is from our mobile application
        \item Device type
        \item Trip length
        \item Whether the trip is during a weekend
        \item The size in km of the diagonal of the smallest rectangular bounds that encapsulates the destination
    \end{enumerate}
The multiclass categorical prediction formulation for location retrieval is much more susceptible to erroneous predictions due to overfitting and out-of-distribution examples. These errors often surface completely irrelevant listings that would break the product experience by displaying a map area that was much larger and irrelevant to the guest's original searched destination. As a result, we only employed a subset of features used in the outgoing retrieval bounds model outlined in \cite{dillontransforming2024} that were the least susceptible to these issues.

\subsection{Location Discretization}
The multiclass categorical prediction formulation for location retrieval requires mapping booked listing locations represented using continuous latitude and longitude coordinates into a tractable discretized categorical space that we can use for multiclass prediction. We employ the S2 Cell system in order to fulfill this requirement.

S2 is a discrete-global-grid system that maps the Earth's surface to the six faces of an inscribed cube, then recursively subdivides each face in a 2-by-2 quadtree pattern. After L levels of subdivision the number of cells is

\begin{equation}
N\!\bigl(L) = 6 \times 4^{(L)}
\end{equation}

The hierarchy is encoded in a 64-bit S2CellId: the top three bits store the parent cube face, a "sentinel 1" bit follows, and 2 bits are added per refinement level (Hilbert-curve ordering guarantees that consecutive cell IDs trace a space-filling path). Cell boundaries are projected quadratic curves, so shapes are spherical quadrilaterals that remain approximately equal-area and never overlap.

This system ensures that location classes are mutually exclusive, deterministic, and similarly sized which is necessary for a performant multiclass classification model. Furthermore, the level-based system provides a systematic and flexible way to tradeoff precision of the S2 cells against tractability of the prediction problem and scalability of retrieval in our production systems.

\subsection{Scalable Extreme Classification}
We chose level 11 S2 cells by choosing the smallest size S2 cell level that we could scalably use for retrieval within our production search retrieval system built using Lucene \cite{apache2license}. This would allow us to maximize the precision of our model within the constraints of our production search system.

There are $\sim$25M level 11 S2 cell ids. This is approximately 100x larger than prediction vocabularies of frontier  large language models \cite{openai2024gpt4}. Furthermore, our training data conditioned on Airbnb bookings is significantly more sparse than all publicly available written language that LLMs use for training.

Given these considerations, training an S2 cell based multiclass classification model is intractable without significant further optimization. We first reduce the scale of the label space by constructing an explicit vocabulary of S2 cells that contain a listing that has been booked on Airbnb. This results in $\sim$360K classes that are necessary for prediction. We then further reduce the scale and model complexity by sharding the training data based on the continent of the searched destination associated with each training example into 3 groups: European Destinations, North + South American Destinations, Non European or American Destinations. This further reduces the complexity from 1 model with $\sim$360K classes to 3 models with $\sim$85K, $\sim$139K, and $\sim$137K classes. These two optimizations allow us to construct models that are supported by modern modeling frameworks and hardware while preserving any cross example generalization that can be learned from training data for similar destinations. Finally, we apply sampled softmax loss \cite{jean2015ssm} with $\sim$25K negative class sampling to efficiently train the set of models. Each model is trained with a learning rate of 0.002 for 16 epochs with early stopping based on cross entropy loss for held out validation data using Tensorflow Horovod distributed training and 2 NVIDIA A100 GPUs and per worker batch size of 45000. Training takes $\sim$3-7 hours depending on the region.

\subsection{Architecture}
We chose deep neural networks among the many possible model architectures that can be applied for classification due to a wide variety of reasons. Primarily, deep neural networks are the only architecture that can scalably and easily support all the challenges and requirements we outlined above regarding extreme classification. Furthermore, it provides more flexibility and freedom with feature engineering and it maintains parity with our existing retrieval bound based location retrieval model \cite{dillontransforming2024} and our suite of ranking models \cite{abdool2020managing} \cite{haldar2019applying} \cite{haldar2020improving}  \cite{tan2023optimizing} \cite{haldar2023learning}. We normalize all continuous features so that they have zero mean and unit variance before training and categorical feature values are mapped to unique embedding vectors that are learned during the optimization process. We employed a 4 layer DNN with layer sizes 1024-2056-1024-256.

\subsection{Retrieval Mechanism}

Given a search request described by feature vector $\mathbf{x}$ (e.g.\ query
destination, guests, nights), the trained multiclass network produces a
probability distribution over the $K=\{85K, 139K, 138K\}$ level-11 S2 cells depending on the continent of the search destination:
\[
\hat{p}_i \;=\; P\!\bigl(C = c_i \mid \mathbf{x};\theta\bigr),
\qquad i = 1,\dots,K .
\tag{1}\label{eq:softmax}
\]

We convert the dense distribution
\eqref{eq:softmax} into a compact set of candidate cells by retaining only
those whose predicted booking likelihood exceeds a hand-tuned threshold
$\lambda\in(0,1)$:
\begin{equation}
\mathcal{R}(\mathbf{x};\lambda) \;=\;
\bigl\{\,c_i \in \mathcal{S}^{(11)} \;\big|\;
      \hat{p}_i \ge \lambda \bigr\}.
\tag{2}\label{eq:relevant-set}
\end{equation}
The choice of $\lambda$ trades off recall (too few cells risk missing viable
inventory) and retrieval cost (too many cells inflate query fan-out).  In
offline tuning, we sweep $\lambda$ and select the value that matches bookings recall of the outgoing retrieval bounds based model.

Each listing document in our search Lucene \cite{apache2license} index has a field
that stores its canonical level-11 S2 cell ID.  Because the
field is single-valued and pre-tokenized, we translate
\eqref{eq:relevant-set} into a disjunctive term query:
\begin{equation}
Q(\mathbf{x}) \;=\;
\bigvee_{c_i \,\in\, \mathcal{R}(\mathbf{x};\lambda)}
\texttt{s2\_lvl\_11}\!:\!c_i .
\tag{3}\label{eq:lucene-or}
\end{equation}
Lucene \cite{apache2license} executes \eqref{eq:lucene-or} yielding the union of listings whose geographic footprint matches
any predicted cell.

\begin{figure}[h]
  \centering
  \includegraphics[width=1.0\linewidth]{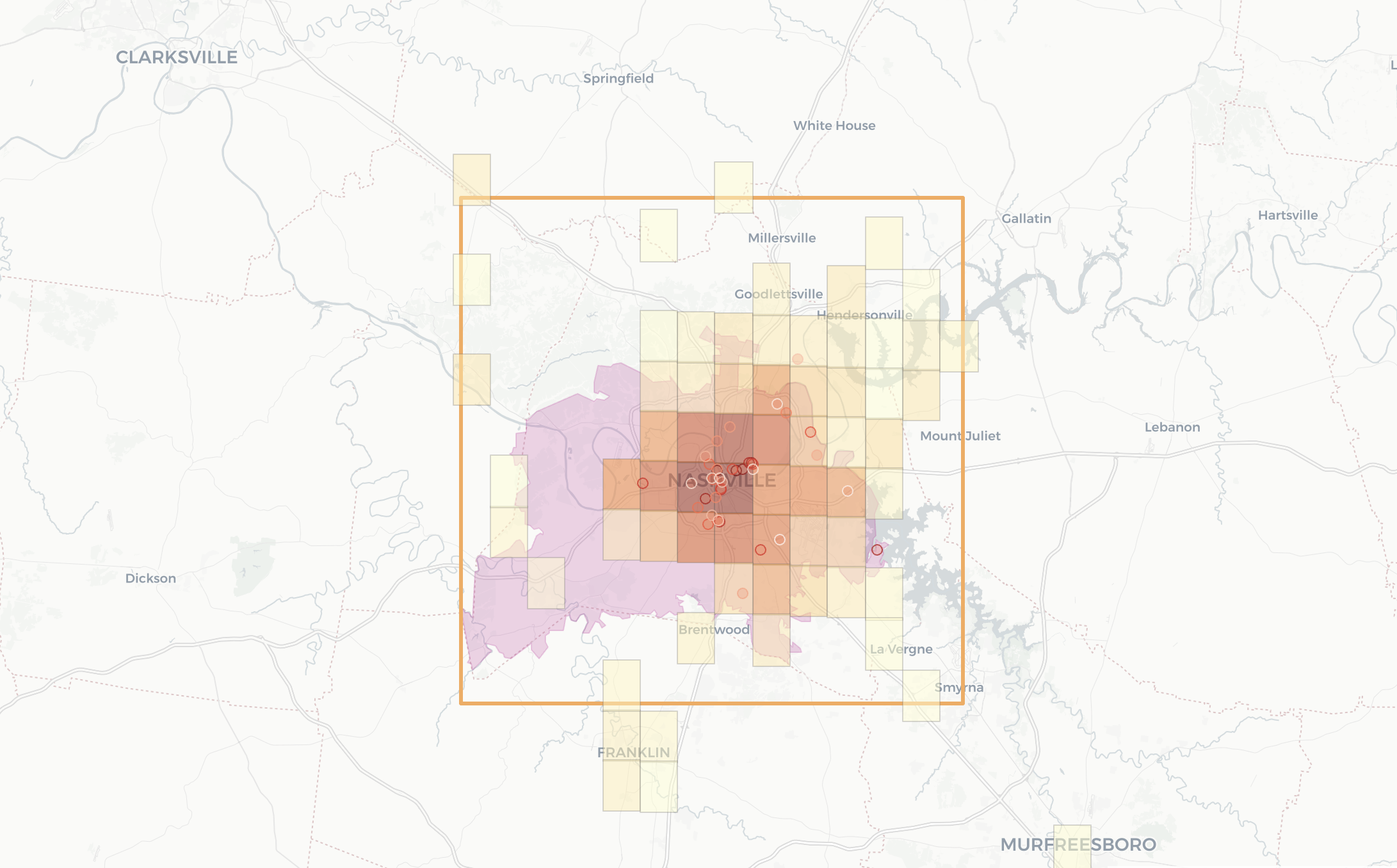}
  \caption{S2 Cell Model Output for Nashville, TN Search with Existing Retrieval Bounds}
  \label{nashville_output}
\end{figure}

Figure \ref{nashville_output} demonstrates S2 cells predictions for a Nashville, TN search after thresholding is applied.

\section{Experiments}
We optimized DNN architectures and hyperparameters based on precision-recall metrics with a held-out dataset. Another important consideration is minimizing negative performance impact because these models will be computed for each search request at serving time. They are much more resource intensive to score with respect to compute, cost, and latency compared to the outgoing retrieval bounds model. Furthermore the retrieval mechanism based on a set of relevant S2 cells is much more computationally expensive than whether a listing is within a rectangular retrieval bounds. Based on these considerations, we choose one model and then run online A/B experiments to measure impact on business metrics such as number of bookings.

\begin{table*}
    \centering
    \resizebox{\textwidth}{!}{%
    \begin{tabular}{l|c|c|c|c|c}
        \hline
         \textbf{Approach}&\textbf{Uncancelled Bookers}&\textbf{Booked Listing Location Recall}&\textbf{Booked Listing Location Precision}&\textbf{Number of Listings Retrieved}\\
        \hline
         S2 Cell Models Against Baseline& +0.17\%& -0.04\%& +11.01\%& +9.91\% \\
    \end{tabular}}
  \caption{Online and Offline Impact of all Approaches}
\vspace{-4mm}
  \footnotesize{Absolute metric values are not shown for company privacy reasons.}
\end{table*}

\subsection{Dataset}
We composed a training dataset composed of 1 years worth of bookings with their attributed searches ($\sim$530M examples) and a held out evaluation dataset of 7 days of bookings with their attributed searches ($\sim$10M examples) from dates immediately after the training range.

\subsection{Metrics}
We consider the following two offline metrics to tune the S2 Cell prediction model and compare it to the baseline retrieval bounds model. The retrieval bounds are converted to a set of S2 cells to accurately compare the two methodologies. Some common classification metrics like AUC are not applicable for our use case because the baseline retrieval bounds model does not have predicted probabilities that can be used for metric computation.

\begin{enumerate}
\vspace{-1mm}
    \item \textbf{Cross Entropy Loss}: Dissimilarity between predicted probability distribution and true probability distribution
    \item \textbf{Booked Listing Location Recall}: Percentage of booked listing location S2 cells that are contained within the predicted positive S2 cells.
    \item \textbf{Booked Listing Location Precision}: Percentage of predicted positive S2 cells that are true positives
    \item \textbf{Number of Listings Retrieved}: The number of search results that pass all retrieval criteria for dates, filters, number of guests, and location retrieval
\end{enumerate}

The primary evaluation to determine whether a new approach is successful is to test whether it has a statistically significant (p-val < 0.05) increase in the following business metric in a massive online A/B experiment.
\\\null \quad \textbf{Uncancelled Bookers}: The unique number of guests that complete a homes booking reservation on Airbnb that did not cancel the reservation during the experiment period. \\

\subsection{Baseline}
The baseline location retrieval method for offline and online evaluation is the retrieval bounds based model trained using the same training data with a weighted piecewise loss. It predicts 4 floats that represent rectangular retrieval bounds \cite{dillontransforming2024}. This model is trained with bookings and searches from the same dates as the S2 Cell models. Its learning rate and model architecture are extensively tuned and it contains features that indicate the region of destinations so there is presumably no difference in performance simply due to the region sharding of the S2 Cell model.

\subsection{Offline Model Selection}
We selected a set of region based S2 Cell models trained with varying learning rates and model complexity that best minimize \textbf{Cross Entropy Loss}. We then select booking probability thresholds $\lambda$ for each region's model by matching the baseline retrieval bound model's \textbf{Booked Listing Location Recall}. This results in the following thresholds for each model:
\begin{enumerate}
    \item European Destination S2 Cell Model:  0.0005
    \item North + South American Destination S2 Cell Model: 0.00075
    \item Non European or American Destinations S2 Cell Model: 0.000625
\end{enumerate}

\subsection{Offline Model Performance}
\begin{figure}[h]
  \centering
  \includegraphics[width=1.0\linewidth]{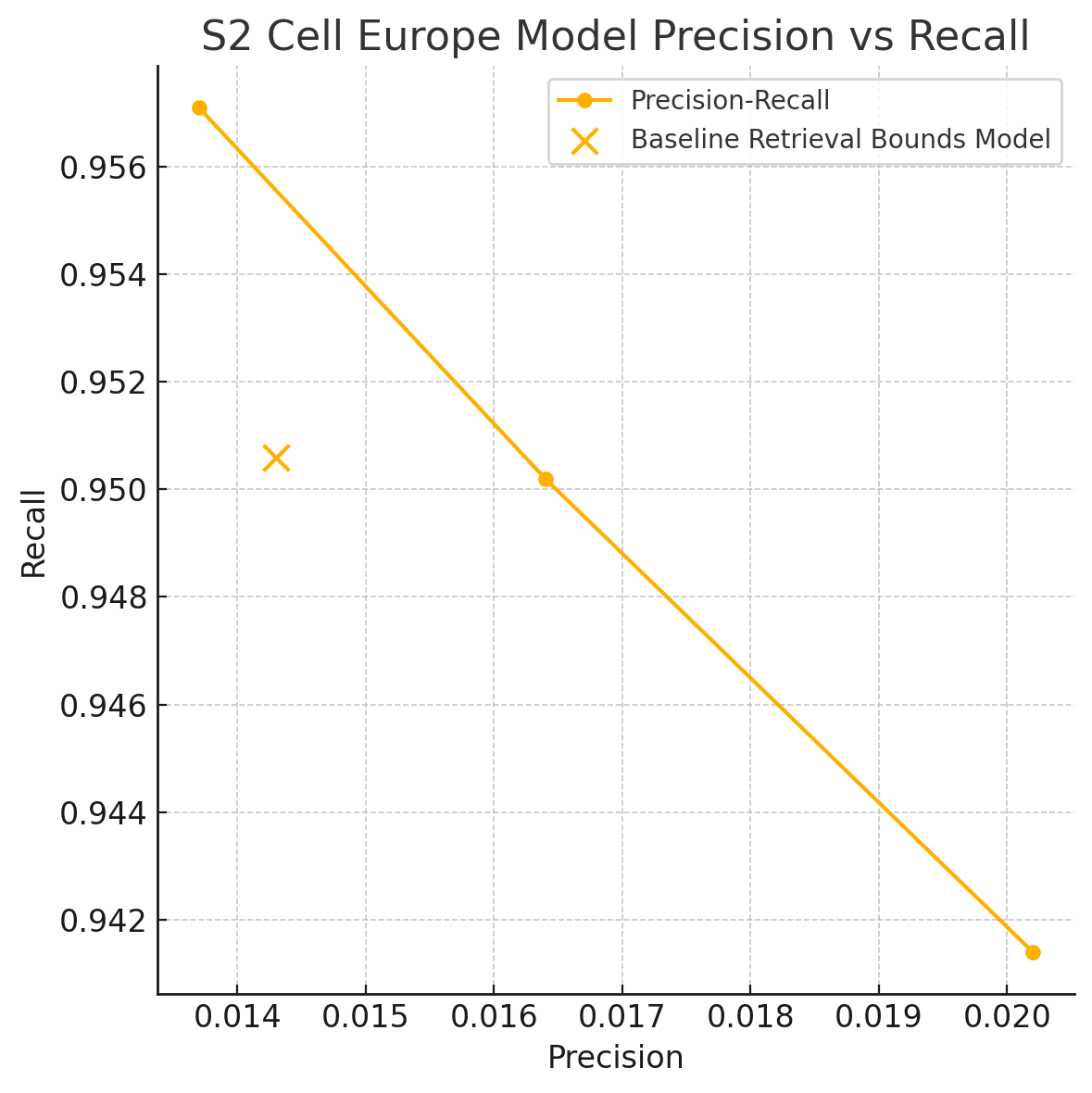}
  \caption{Europe S2 Cell Model Precision vs Recall}
  \label{europe_perf}
\end{figure}

\begin{figure}[h]
  \centering
  \includegraphics[width=1.0\linewidth]{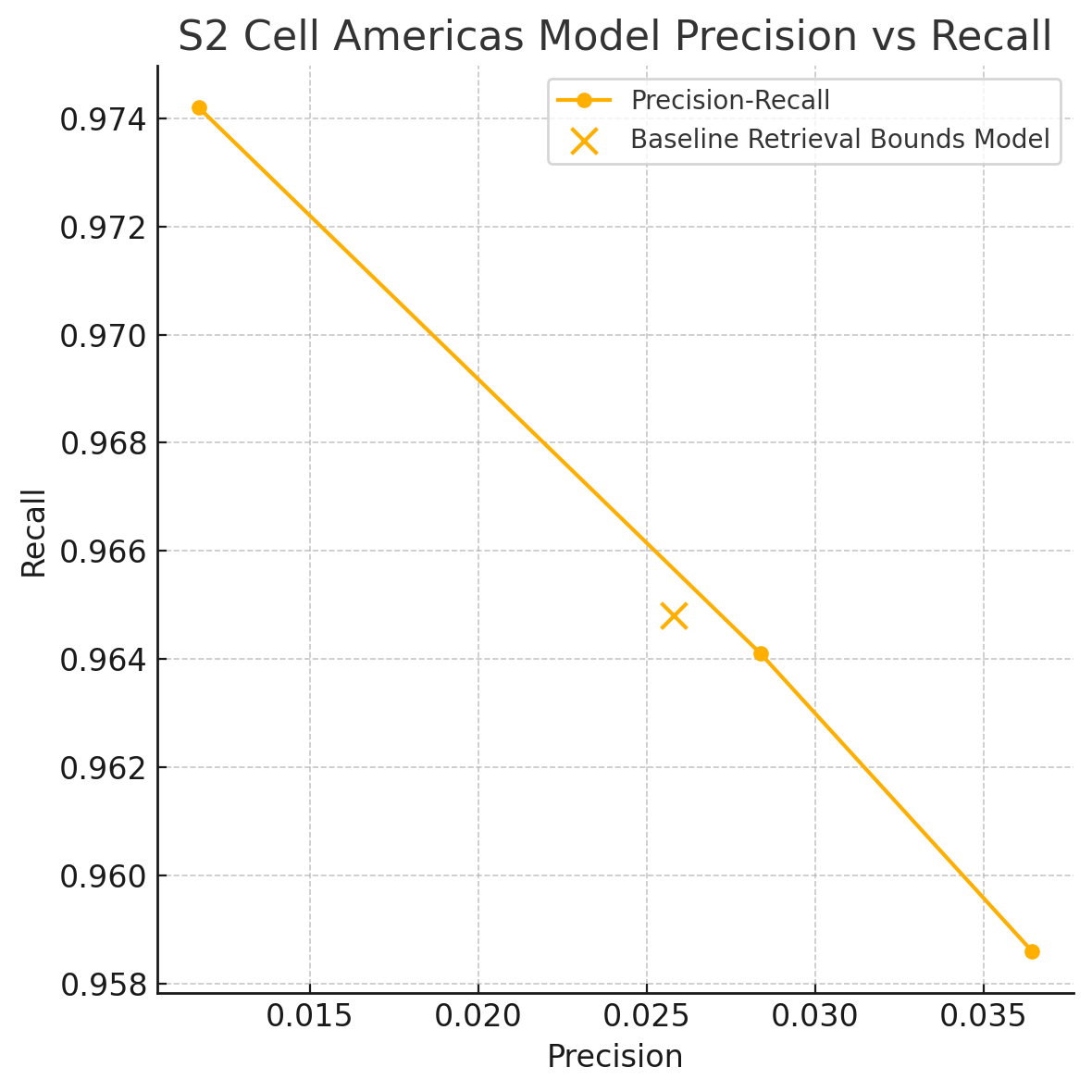}
  \caption{Americas S2 Cell Model Precision vs Recall}
  \label{america_perf}
\end{figure}

\begin{figure}[h]
  \centering
  \includegraphics[width=1.0\linewidth]{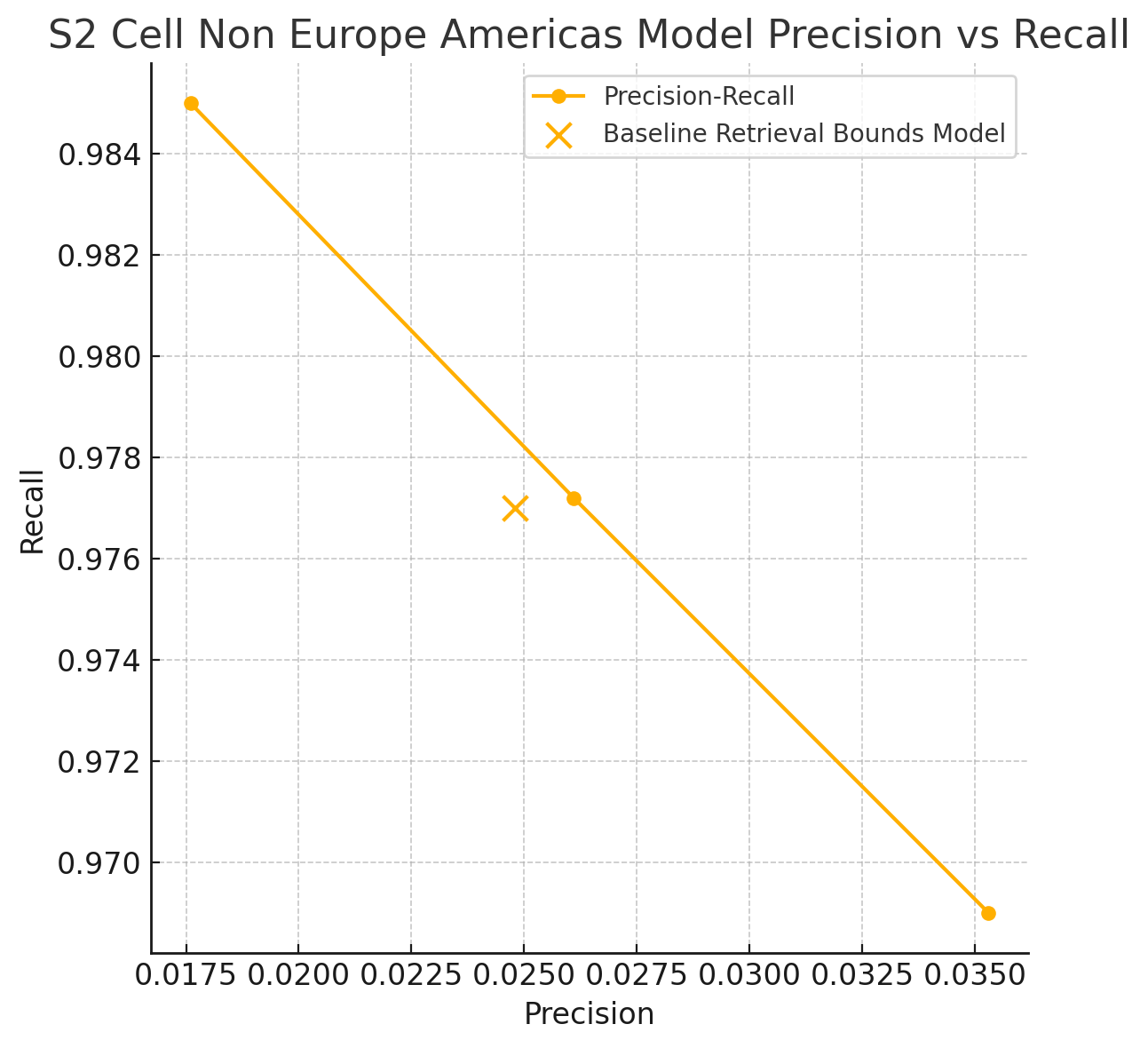}
  \caption{Non European-American S2 Cell Model Precision vs Recall}
  \label{asia_perf}
\end{figure}

All three region based models perform better with respect to precision-recall tradeoffs compared to the outgoing retrieval bounds based methods.
They achieve approximately the same \textbf{Booked Listing Location Recall} (-0.04\%) while having +11.01\% higher \textbf{Booked Listing Location Precision}. Performance is shown in \ref{europe_perf}, \ref{america_perf}, \ref{asia_perf}. Despite similar recall and higher precision, the three region based models increase \textbf{Number of Listings Retrieved} by 9.91\% providing guests with more inventory in bookable locations.

\subsection{Online Model Performance}
The three region based S2 Cell models were tested against the baseline model described above in a combined online A/B experiment including $\sim$66M users. The test showed that the S2 Cell models increased \textbf{Bookers that did not cancel} by \textbf{+0.17\%}.

\section{Conclusion}
We hypothesized that a guest's search context contains enough signal to
localize the eventual booking into a small set of level-11 S2 cells.  Treating
each cell as a class, we trained an extreme-multiclass model that outputs
posterior probabilities over $\mathcal{S}^{(11)}$.  By thresholding these
posteriors and OR-querying the corresponding cell IDs in a Lucene \cite{apache2license} index, we
replace rectangular retrieval bounds retrieval with high precision categorical location retrieval.

Offline, the method matches the legacy retrieval bounds model's recall, lifts precision by +11.01\%, and increases inventory by 9.91\%; online, it delivers a +0.17 \% gain in uncancelled bookers.

Next steps include learning the threshold per query, exploiting the S2
hierarchy for hierarchical soft-max, and improved feature engineering. More broadly, any marketplace that retrieves geo-constrained supply can adopt this classify-then-query pattern to improve recommendation performance without rewriting its search stack.

\bibliographystyle{ACM-Reference-Format}
\bibliography{base}

\end{document}